\documentclass[%
 reprint,
 amsmath,amssymb,
 aps,
prl,
]{revtex4-1}

\usepackage{graphicx}
\usepackage{dcolumn}
\usepackage{bm}

\begin{document}

\preprint{APS/123-QED}

\title{Janus and Huygens' dipolar sources for near-field directionality}






\author{Michela F. Picardi}
\email{michela.picardi@kcl.ac.uk}
\author{Anatoly V. Zayats}
\author{Francisco J. Rodr\'{i}guez-Fortu\~{n}o}%
\affiliation{Department of Physics, King's College London \\
Strand, London, WC2R 2LS, United Kingdom}%

\date{\today}

\begin{abstract}

Controlling directionality of emission, scattering and waveguiding is an important requirement in quantum optical technology, integrated photonics and new metasurface designs, as well as radio and microwave engineering. Recently, several approaches have been developed to achieve unidirectional scattering in the far-field relying on Huygens' dipolar sources, and in waveguided optics based on spin-Hall effects involving circularly polarised electric or magnetic dipoles, all of which can be realised with plasmonic or dielectric nanoparticles. Here we show that there exists a dipolar source complimentary to Huygens' dipole, termed Janus dipole, which is not directional in the far-field, but its coupling to waveguided modes is topologically protected so that it is allowed on one side of the dipole but not on the opposite side. The near field directionality of the Huygens' dipole is also revealed and a generalised Kerker's condition for far- and near-field directionality is introduced. Circular electric and magnetic dipoles, together with Huygens' and Janus dipolar sources, form a complete set of directional dipolar sources in far- and near-field, paving the way for promising applications.


\end{abstract}

\pacs{Valid PACS appear here}
\maketitle




Nanoscale emitters, scatterers and their assemblies have been recently considered for scalable photonic circuitry, where the requirements on miniaturization and efficient coupling to photonic modes are strict, metasurface designs enabling flat lenses and hologrammes, as well as quantum optical technologies \cite{Photonics2014,yu2014flat,chang2014quantum}. They can be realised as strongly resonant plasmonic or high-index dielectric nanoparticles supporting electric and/or magnetic dipolar resonances. Going beyond linearly polarised dipoles opens unexpected opportunities for electromagnetic designs. Near field interference and related directional excitation of fields from circularly polarized electric and magnetic dipoles \cite{Rodriguez-Fortuno2013,Kapitanova2014,Aiello2015,Espinosa-Soria2016,Mechelen2015,Coles2016,LeFeber2015,zharov2016control,wang2017photonic,garoli2017helicity,picardi2017unidirectional} have fascinating applications in quantum optics \cite{marrucci2015quantum,Luxmoore2013,mitsch2014quantum}  and in novel nanophotonic devices such as nanorouters, polarimeters, and non-reciprocal optical components \cite{Petersen2014,Neugebauer2014,espinosa2017chip, OConnor2014,rodriguez2014universal,rodríguez2014resolving,rodriguez2014sorting,sayrin2015nanophotonic,ma2016hybrid}. These effects rely on the photonic spin-Hall effect exploiting the phenomenon of spin-momentum locking in evanescent and guided waves \cite{bliokh2015quantum,bliokh2012transverse,Junge2013,bliokh2014extraordinary}: in essence, the spin of the dipole can be matched to the spin of the guided fields to be directionally excited. While electromagnetic spin is a quantity that accounts for the relative amplitude and phase of the different electric field \emph{or} magnetic field  components of a guided wave --describing the rotation of these two vectors $\mathbf{E}$ and $\mathbf{H}$-- spin does not account for the relative amplitude and phases \emph{between} electric and magnetic components. Engineering superpositions of electric and magnetic dipoles and their interference \cite{sinev2017dielectric,lee2012role,evlyukhin2015resonant,fu2012directional} takes care of this limitation.  


%
%
%
%
%
%
%

There is a well-known dipolar source which explicitly exploits these relations to achieve \emph{far-field} directionality: the Huygens' antenna. This source combines two orthogonal linearly polarized electric $p$ and magnetic $m$ dipoles (Fig. 1) satisfying Kerker's condition \cite{kerker1983electromagnetic,zambrana2013duality}:
\begin{equation}\label{eq:kerker}
p = \frac{m}{c} ,
\end{equation}
with $c$ being the speed of light. The radiation diagram of such an antenna is highly directional and has zero back-scattering, due to the interference of magnetic and electric dipole radiation. 
These antennas are attracting great attention due to the feasibility of implementing them using high-index dielectric nanoparticles \cite{evlyukhin2012demonstration, permyakov2015probing,kuznetsov2016optically}, with applications in null back-scattering, metasurfaces, and all-dielectric mirrors \cite{geffrin2012magnetic,person2013demonstration,staude2013tailoring,nieto2011angle,coenen2014directional,wei2017adding,paniagua2016generalized}. 

Here we show that Huygens' sources can be generalized to achieve near-field directionality, and that there exists a dipolar source complementary to a Huygen's dipole, which we term Janus dipole, with a different relation between the phases of electric and magnetic dipoles, which is not directional in the far-field, but has unique near-field properties allowing ``side"-dependent coupling to guided modes. Together, Huygens', Janus, circular electric and magnetic dipoles (as well as the infinite spectrum of their linear combinations) provide a general closed solution to dipolar far- and near-field directionality that takes into account the topology of the vector structure of free space and guided electromagnetic fields. These dipolar sources can be experimentally realised as plasmonic, dielectric and hybrid nanoparticles.

\begin{figure}
\includegraphics[width=\linewidth]{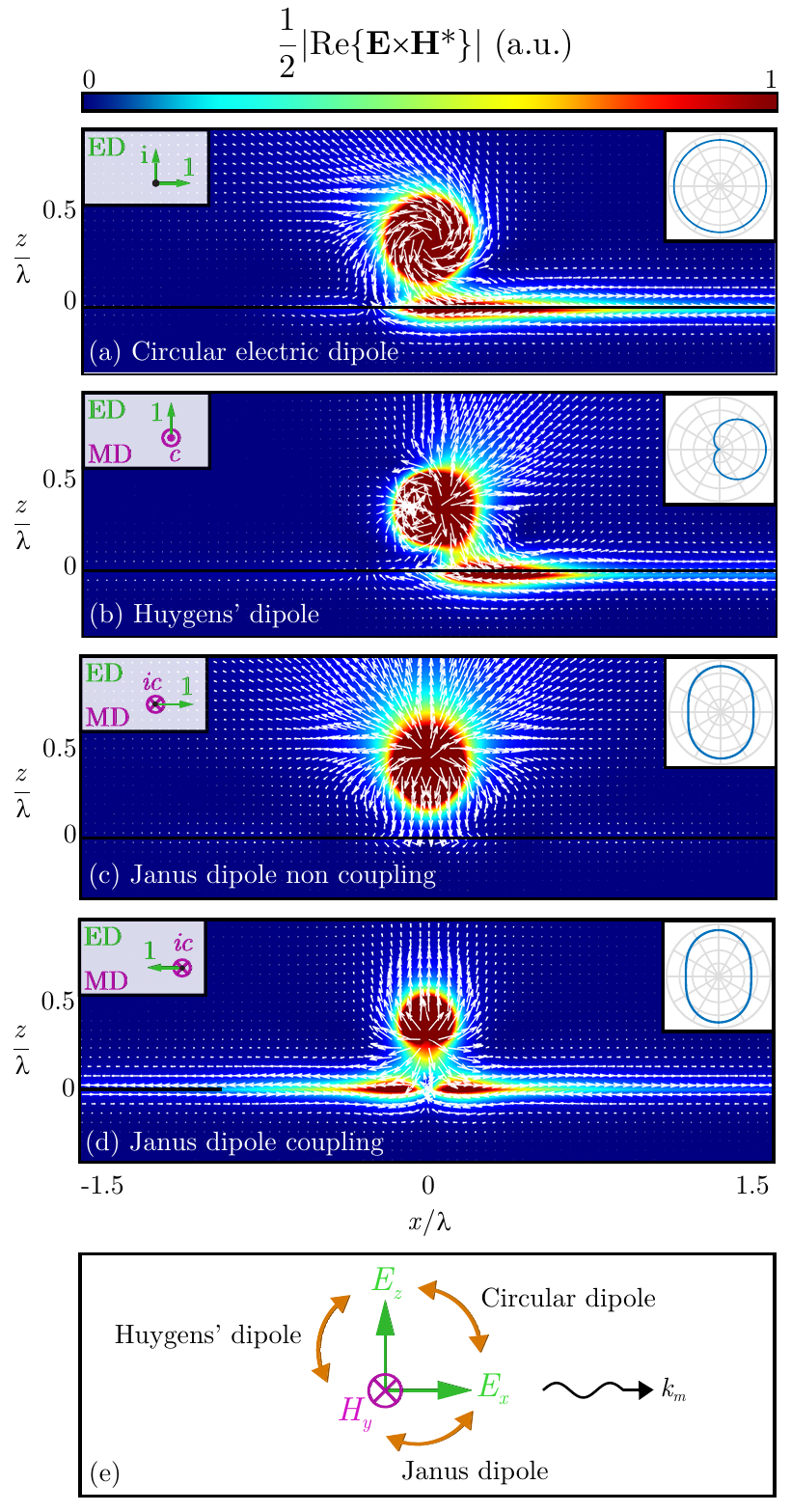}
\caption{ Power flow induced by (a) a circularly polarized electric dipole $\mathbf{p}=(1,0,i)$, $\mathbf{m}=(0,0,0)$; (b) a Huygens' antenna $\mathbf{p}=(0,0,1)$, $\mathbf{m}=(0,-c,0)$; (c,d) a Janus dipole $\mathbf{p}=(\pm 1,0,0)$, $\mathbf{m}=(0,ic,0)$ in non-coupling (c) and coupling (d) orientation, in close proximity ($z_0 = 0.3 \lambda$) to an interface of a material with $\varepsilon= -1.5$ and $\mu=1$. Calculated by integration of the angular spectra of the dipole field. (e) Schematic of field components excited by each source. The insets show the orientation of the dipoles and the far field radiation diagrams.\label{fig:1}}
\end{figure}

In order to illustrate the properties of the considered dipolar sources, Figure \ref{fig:1} shows the time averaged power flow vector generated by (a) a circular dipole, (b) a Huygens' antenna, and (c,d) a Janus dipole for its two orientations, all placed over a waveguiding surface. We used metallic surfaces supporting surface plasmons as simple examples, but the directionality of the dipoles is universal and completely independent of the waveguide's nature. As can be seen, the first two sources lead to directional evanescent wave excitation of guided modes along the waveguide. While this is known for circular dipoles \cite{Rodriguez-Fortuno2013,Kapitanova2014,Aiello2015,Espinosa-Soria2016,Mechelen2015,Coles2016,LeFeber2015,zharov2016control,wang2017photonic,garoli2017helicity,Junge2013,picardi2017unidirectional}, Huygens' antennas have been massively studied for their strong directional radiation diagram, but their near field directionality had not been explored. The direction of excitation of these sources can be switched by flipping the sign of one of their two dipole components, which can be experimentally achieved tuning polarization and wavelength of the light illuminating the nanoparticle, with respect to its electric and magnetic resonances.

The Janus dipole has an intriguing property: it either shows (c) a complete absence of coupling, in which it does not excite waveguide modes at all or (d) excitation of the guided mode in both directions. This is determined by which `side' of the dipole is facing the waveguide. Inverting the sign of one component in the Janus dipole will change the side facing the waveguide, like when flipping a coin, and this will switch the coupling on and off [Figs.~\ref{fig:1}(c,d)]. 

An intuitive explanation of the three sources can be obtained as follows. Fermi's golden rule \cite{Aiello2015,Luxmoore2013,Coles2016,LeFeber2015,marrucci2015quantum,Mechelen2015,Espinosa-Soria2016} dictates that the coupling efficiency between an electric $\mathbf{p}$ and magnetic $\mathbf{m}$ dipole source and a waveguide mode is proportional to $|\mathbf{p}\cdot\mathbf{E}^{*}+\mathbf{m}\cdot\mu \mathbf{H}^{*}|^2$  where $\mathbf{E}$ and $\mathbf{H}$ are the electric and magnetic fields of the mode calculated at the location of the dipoles, and $\mu$ is the permeability of the medium. In Fig.~\ref{fig:1}, the dipoles are interacting with a $p$-polarized waveguide mode, so the only non-zero field components are the transverse electric and magnetic fields $E_z$ and $H_y$, and the longitudinal field $E_x$. These are shown in Fig.~\ref{fig:1}(e). The circular dipole uses the dipole components $p_x$ and $p_z$ to couple with the $E_x$ and $E_z$ components of the mode. Its directional behaviour relies on the well understood spin-momentum locking between these components, which dictates that $E_x$ and $E_z$ undergo a fixed amplitude and phase relationship, resulting in $\mathbf{E}$ having a circular polarization, associated with a transverse spin, whose sense of rotation depends on the propagation direction \cite{Junge2013,Aiello2015,Mechelen2015,bliokh2012transverse,bliokh2014extraordinary,bliokh2015quantum}. The dipole exploits this such that $\mathbf{p} \cdot \mathbf{E}^* = 0$ for the mode propagating to the left or right, thereby showing unidirectional excitation in the opposite direction. Analogously, circular magnetic dipoles can directionally excite $s$-polarized modes by exploiting $\mathbf{m} \cdot \mu \mathbf{H}^* = 0$. Both are possible thanks to the longitudinal component of the evanescent fields.

To describe the nature of the other two sources, however, we must also take into account the relative phase and amplitude between the $\mathbf{E}$ and $\mathbf{H}$ components, not usually considered in spin-direction locking. Their relation can be exploited such that the electric  and magnetic coupling terms interfere destructively between each other $\mathbf{p} \cdot \mathbf{E}^* + \mathbf{m} \cdot \mu \mathbf{H}^* = 0$. In other words, the mode excited by the electric dipole $\mathbf{p}$ in a given direction is exactly cancelled out by the one excited by the magnetic dipole $\mathbf{m}$ after their superposition. The Huygens' source exploits the fixed relative amplitude and phase that exists between the transverse field components $E_z$ and $H_y$, which depends on the propagation direction of the mode. This relation is a well-known property of plane waves which extends directly into evanescent and guided waves.

On the other hand, there is another pair of components that we can consider [Fig.~\ref{fig:1}(e)]. The Janus dipole exploits the locked amplitude and phase relation that exists between $H_y$ and the longitudinal electric field $E_x$. The unique feature of the Janus dipole, which distinguishes it from the other two, is that this interference can be achieved simultaneously in both propagation directions of a mode, because the ratio between $E_x$ and $H_y$ is independent of the mode's left or right propagation direction. This is a remarkable topological feature of evanescent wave polarization in addition to transverse spin \cite{bliokh2015quantum}. It enables us to design an electric dipole $p_x$ and magnetic dipole $m_y$ such that their mode excitations interfere destructively in \emph{both} directions. Note that the independence of the ratio between longitudinal electric field and transverse magnetic field with respect to the propagation direction (time reversal) is universally true, at any location, on any translationally invariant waveguide. This can be proven by considering a mirror reflection on a plane perpendicular to the waveguide axis: the propagation direction changes, but the ratio of the two components does not, due to the simultaneous flipping of both the longitudinal component and the magnetic component, which being a pseudo-vector changes sign under reflections. Thus, a Janus dipole can be designed to achieve polarization and position-dependent ``non-coupling'' in any scenario where longitudinal fields are present, such as inside nanowires and photonic crystal waveguides, not being limited to evanescent coupling as illustrated here.

Both the circular and Janus dipole involve the longitudinal component of the field, while the Huygens' source does not. This explains why circular and Janus dipoles are not directional in the far field \cite{Rodriguez-Fortuno2013,picardi2017unidirectional}, as plane waves have no longitudinal field. The Huygens' source, instead, is always directional, as it exploits the relation between the transverse fields which exists in plane waves and evanescent waves alike. Another crucial difference lies in the intrinsic symmetry of the sources themselves. The circular dipole has rotational symmetry around the $y$ axis. The Huygens' source --recalling that the magnetic moment is a pseudo-vector-- is mirror symmetric with respect to the $z = 0$ plane, while the Janus source is mirror symmetric with respect to the $x = 0$ one. This difference in symmetry leads to a remarkable result when considering coupling of the dipoles surrounded by waveguides.
 
\begin{figure}[htbp]
\includegraphics[width=\linewidth]{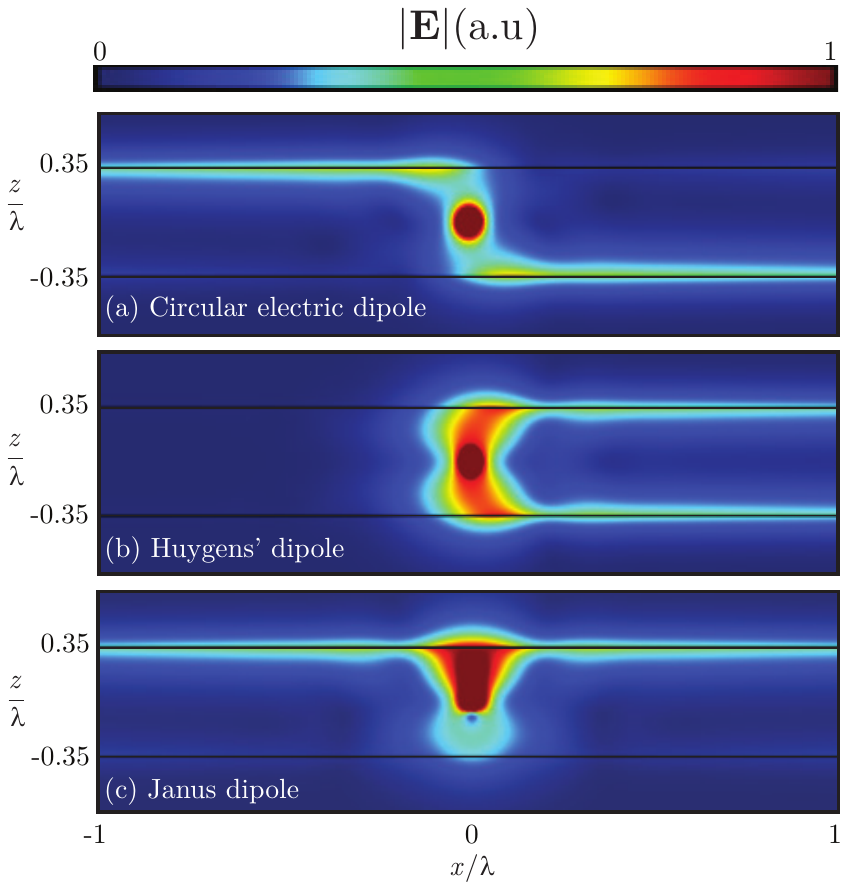}
\caption{Amplitude of the electric field generated by (a) a circular dipole, (b) a Huygens' antenna and (c) a Janus dipole embedded in the centre of a metal-air-metal waveguide, with $\varepsilon= -1.5 + 0.02i$ and $\mu=1$. The distance between the two waveguides is $0.7\lambda$.\label{fig:2}}
\end{figure}

Figure~\ref{fig:2} shows the three dipole sources embedded between two waveguides, again metallic surfaces only as an example, at a distance such that light from the dipole can couple to both waveguides, but with negligible coupling between the waveguides for the propagation distances considered. The circular dipole couples into opposite directions for the waveguides placed above or below the dipole, while the Huygens' dipole couples in the same direction for both. Most interestingly, while these two sources exhibit left-right directionality, the Janus dipole exhibits a front-back directionality. While it does not excite the waveguide placed below, it does however excite both directions in the waveguide above it, regardless of its distance to either. In this way, the Janus dipole is topologically protected from coupling into the waveguide facing its non-coupling side. This arises because the ratio between $E_x$ and $H_y$ in evanescent waves is independent of the propagation direction but does depend on the \emph{direction of evanescent decay}. This remarkable and inherently broadband behaviour suggests novel potential applications in optical nanorouting and signal processing. Importantly, all the directionality properties described in Fig.~\ref{fig:2} are robust and independent of the distance of the dipoles to the waveguides. The symmetry of excitations follows directly from that of the sources themselves.

The design of dipoles exhibiting near-field interference can be done in a general case by means of the Fermi golden rule, as long as the modal fields are known. However, we can provide simple expressions for the specific case of dipoles coupling into the evanescent fields of a planar waveguide extending along the transverse $(x, y)$ plane (Figs.~\ref{fig:1} and \ref{fig:2}). For simplicity, we can align our reference system with the propagation direction of the mode, such that the wave-vector of the evanescent field is given by $\mathbf{k} = (k_x,k_y,k_z) = (\pm k_m, 0, \pm i \alpha_m)$, where $k_m$ is the propagation constant of the mode, $\alpha_m = (k_m^2 - k^2)^{1/2}$ accounts for the evanescent nature and $k$ is the wave-number of the medium. The sign of $\pm k_m$ accounts for the direction of propagation, while the sign of $\pm i \alpha_m$ depends on the direction of evanescent decay, which depends on whether the waveguide is below (positive) or above (negative) the dipole. We can write the three components of $p$-polarized modes in a vector of the form $\mathbf{F}_p = (E_x,c \mu H_y, E_z)$ and the corresponding dipole moment components as a vector $\mathbf{q}_p = (p_x,m_y/c,p_z)$ so that the Fermi's golden rule is reduced to a simple scalar product $\left| \mathbf{q}_p \cdot \mathbf{F}^*_p \right| ^2$. Maxwell's equations demand that $p$-polarized fields with $k_y = 0$ are given by $\mathbf{F}_p \propto (\frac{\pm i \alpha_m}{k}, 1, -\frac{\pm k_m}{k})$ \cite{picardi2017unidirectional}, irrespective of the nature of the waveguide. The single key aspect underpinning all phenomena described in this work is that each pair of these three components has a fixed amplitude and phase relation between them. Indeed, each of the three elemental dipole sources is derived from the relationship between each of the three possible pairs of field components [Fig.~\ref{fig:1}(e)]. Notice that the ratio between $E_x$ and $E_z$ depends on both the direction of propagation $\pm k_m$ and on the sign of the evanescent decay (evanescent field gradient) $\pm i \alpha_m$, as known for spin-direction locking. The ratio between $E_z$ and $H_y$ depends only on propagation direction, thus explaining Huygens' properties, while the ratio between $E_x$ and $H_y$ is independent of the propagation direction, as proved earlier in a general case, but depends on the sign of $\pm i \alpha_m$, explaining the unique behaviour of the Janus dipole.

To obtain near-field interference effects, we can solve the equation that achieves zero coupling of the dipoles into a given mode:
\begin{equation}\label{eq:e_zero}
\mathbf{q}_p \cdot \mathbf{F}_p^* = \left(p_x,\frac{m_y}{c},p_z\right) \cdot \left(\frac{\pm i \alpha_m}{k}, 1, -\frac{\pm k_m}{k}\right)^* = 0.
\end{equation}
Mathematically, this simple equation defines a geometric plane of solutions given by the sub-space of dipole vectors $\mathbf{q}_p$ which are orthogonal to $\mathbf{F}_p$, and provides a unified view of all the possible ways to achieve directionality of $p$-polarized modes when using any electric and magnetic dipole source. The fixed relationships between the field components translate directly into conditions between the dipole components. Each of the three sources discussed above correspond to intersections of this plane with the $p_x$, $m_y$, or $p_z = 0$ planes. Alternatively, each dipole corresponds to the intersection of two planes given by Eq.~\ref{eq:e_zero} but for different pairs of sign combinations in $k_m$ and $\alpha_m$, explaining why each case shows zero excitation of exactly two directions in Fig.~\ref{fig:2}. A compact summary of the mathematical solutions to this equation is given in Table \ref{tab:table1}. Notice that the dipoles are fine-tuned to achieve a perfect contrast ratio for a specific mode $k_m$, but the non-optimized versions, in which $(p_x,m_y/c,p_z) \propto (1,0,\pm i)$, $(0,\pm 1, 1)$ and $(1,\pm i,0)$, also work remarkably well as shown in Fig.~\ref{fig:1}. Each of the three elemental sources corresponds to a vector within the same plane of solutions, so that each is obtainable as a linear superposition of the other two. Finally, we can consider the entire geometric plane of solutions obtained by linear combinations of the elemental sources, resulting in an infinite range of electric and magnetic dipoles that verify Eq.~\ref{eq:e_zero}. 

\begingroup
\begin{table}
\begin{center}
\caption{Elemental dipole sources for near-field directionality in the $(x,z)$ plane. Optimized dipoles use $\hat{k}_m=k_m^*/k$ and $\hat{\alpha}_m=\alpha_m^*/k$, while non-optimized dipoles use $\hat{\alpha}_m, \hat{k}_m\approx 1$ and also show good performance. The sign of $\pm \hat{k}_m$ will determine whether we nullify the mode coupling to the right or to the left, respectively, while the sign of $\pm \hat{\alpha}_m$ will determine whether the waveguide is below or above the dipole, respectively. In the general solution, $\mathbf{q}_{p/s}^{i}$ and $\mathbf{q}_{p/s}^{j}$ stand for any two of the three elemental dipoles and $a, b$ are arbitrary complex coefficients.} \label{tab:table1}

\begin{tabular}{ |c|c|c|c| } 
\hline
  & $p$-polarization & $s$-polarization \\
  & $\mathbf{q}_p = (p_x,m_y/c,p_z)$ & $\mathbf{q}_s = (m_x/c,p_y,m_z/c)$ \\
\hline
Elliptical & $(\pm \hat{k}_m,0,\mp i \hat{\alpha}_m)$ & $(\pm \hat{k}_m,0,\mp i \hat{\alpha}_m)$ \\ 
Huygens & $(0,\pm \hat{k}_m,1)$ & $(0,\pm \hat{k}_m,-1)$ \\ 
Janus & $(1,\pm i \hat{\alpha}_m,0)$ & $(-1,\pm i \hat{\alpha}_m,0)$ \\ 
\hline
General & $\mathbf{q}_p = a \mathbf{q}_p^{i} + b \mathbf{q}_p^{j}$ & $\mathbf{q}_s = a \mathbf{q}_s^{i} + b \mathbf{q}_s^{j}$ \\ 
\hline
\end{tabular}
\end{center}
\end{table}
\endgroup

Analogous considerations are valid for $s-$ polarized modes. In this case, Maxwell's equations imply that the electromagnetic fields when $k_y = 0$ are given by $\mathbf{F}_s = (c \mu H_x, E_y, c\mu H_z) \propto (\pm i \frac{\alpha_m}{k}, -1, -\frac{\pm k_m}{k})$, and writing the relevant dipole components as $\mathbf{q}_s = (m_x/c,p_y,m_z/c)$, the near-field destructive interference condition based on Fermi's golden rule can be written as $\mathbf{q}_s \cdot \mathbf{F}_s^* = 0$. Solutions are given in Table \ref{tab:table1}. In complete physical analogy to the $p$-polarized case, the same three elemental dipoles can be derived, but swapping the roles of the electric and magnetic moments.

\begin{figure}[htbp]
\includegraphics[width=1\linewidth]{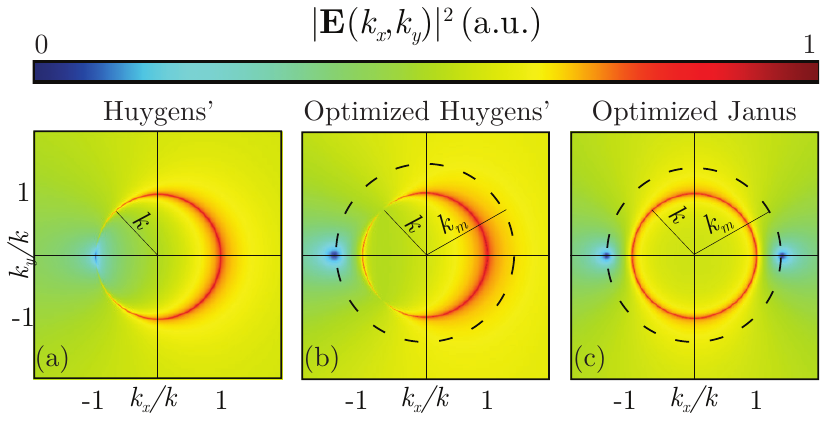}
\caption{Angular spectra of (a,b) Huygens' antennas satisfying (a) usual Kerker's condition (Eq.~\ref{eq:kerker}) and (b) optimized Kerker's condition (Eq.~\ref{eq:opt_kerker}); (c) a Janus, for a waveguide with $k_m = \sqrt{2}k$.\label{fig:3}}
\end{figure}

The optimized conditions for the Huygens' and Janus dipoles can be written compactly as:
\begin{equation}
\label{eq:opt_kerker}
\frac{\pm k_{m}^*}{k}p  = \frac{m}{c} \qquad\text{and}\qquad \frac{\pm i \alpha_{m}^*}{k}p  = \frac{m}{c}.
\end{equation}
\noindent The first equation corresponds to the Huygens' dipole and constitutes a generalized Kerker's condition which works for both evanescent and propagating waves. It, in fact, reduces to the \textit{usual} Kerker's condition (Eq.~\ref{eq:kerker}) for $k_{m}=k$.  The angular spectrum of the source \cite{supplementarycite} provides a convincing visual explanation of this optimization. In Fig.~\ref{fig:3} we plot the angular spectrum of the fields below a Huygens' antenna in two different cases. Panel (a) corresponds to the usual Kerker's condition (Eq.~\ref{eq:kerker}), and it can be seen that the spectrum is zero on the transverse wave-vector $\mathbf{k}_t=(-k,0)$, lying exactly on the light cone. In panel (b) we show the spectra of the generalized Huygens' dipole (Eq.~\ref{eq:opt_kerker}), and we see that it is zero at $\mathbf{k}_t=(-k_{m},0)$, corresponding to the mode supported by the waveguide. Comparing the amplitude of both spectra at $(\pm k_{m},0)$, both cases will excite modes preferentially in the direction $+k_{m}$, but only the optimized Kerker condition will achieve a perfect contrast ratio. The second Eq.~\ref{eq:opt_kerker} corresponds to the Janus dipole and, in the limit $k_m \to \sqrt{2} k$ reduces to $\pm i p = m/c$. It greatly resembles Kerker's condition but with a phase difference. The angular spectrum of the Janus dipole is shown in Fig.~\ref{fig:3}(c) and shows a null spectral component for both $\mathbf{k}_t = (k_m,0)$ and $(-k_m,0)$. Thus from simple momentum matching arguments, the dipole itself is incapable of coupling into the modes of the waveguide at all because it lacks the required angular components. While the above considerations have been derived for a planar waveguide, we would like to emphasize that following a spectral interpretation, 
all dipoles derived in Table \ref{tab:table1} are excellent approximations to their optimum when placed near arbitrary waveguides, as was shown in Ref.~\cite{picardi2017unidirectional} based on momentum conservation arguments.

\begin{figure}
\includegraphics[width=\linewidth]{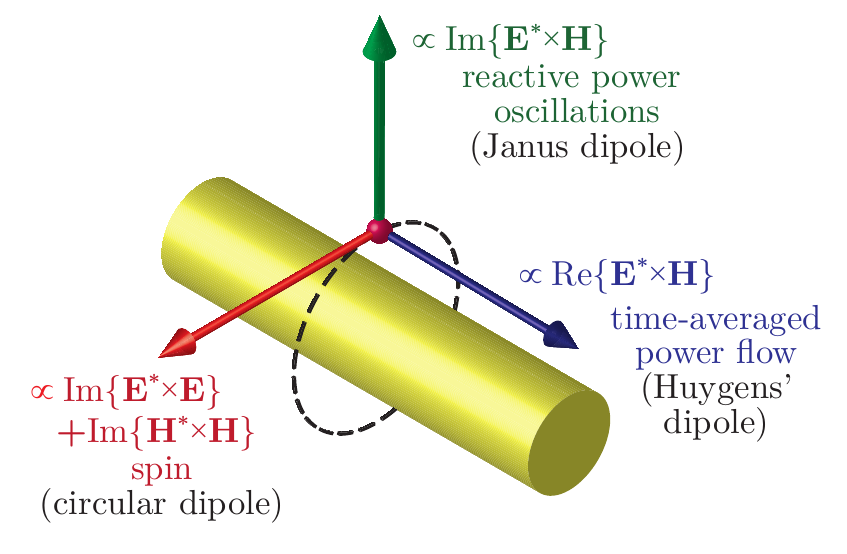}
\caption{Schematic depicting a triad of vectors for a guided mode: time-averaged power flow, reactive power and spin vector. Each closely related to one of the three sources.\label{fig:vectors}}
\end{figure}

Further insight can be obtained using well-known electromagnetic quantities (Fig.~\ref{fig:vectors}). The Huygens' source is often explained in terms of the time-averaged Poynting vector $\propto \mathrm{Re}\left[\mathbf{E}^*\times\mathbf{H}\right]$; it uses orthogonal electric and magnetic dipoles, in phase, to produce the corresponding fields associated with a net power flow in a given direction. The canonical spin angular momentum $\propto \mathrm{Im}\left[\mathbf{E}^*\times\mathbf{E}+\mathbf{H}^*\times\mathbf{H}\right]$ accounts for the spin of vectors $\mathbf{E}$ and $\mathbf{H}$, arising when either field has orthogonal components phase-shifted by $\pi/2$, exactly as generated by circular dipoles. The Janus dipole is associated with the vector $\propto \mathrm{Im}\left[\mathbf{E}^*\times\mathbf{H}\right]$. This expression resembles that of spin, but mixing electric and magnetic components. It arises when $\mathbf{E}$ and $\mathbf{H}$ are orthogonal and $\pi/2$ out of phase, as produced by the Janus dipole. Notice that this vector is the imaginary part of the complex Poynting vector, known to signify the direction of \emph{reactive power} in which harmonic oscillations of power occur with no net flow. The vector points in the direction of evanescent decay. The Janus dipole can thus match or oppose these oscillations. The three vector quantities, each associated with one of the sources, are known to form a locked triad at each point near a waveguide \cite{Mechelen2015}, as shown in Fig.~\ref{fig:vectors}, accounting for the sources' symmetries and behaviour.


Previous approaches to guided optics directionality from dipolar sources made use of the spin of the guided mode's fields $\mathbf{E}$ and $\mathbf{H}$, neglecting their mutual amplitude and phase relations. By considering the whole vector structure of electromagnetic fields we provide a unified theory describing all possible dipole sources exhibiting far- and near-field directionality, opening the way to entirely new types of directional coupling. The implementation of these new sources using resonant plasmonic or dielectric nanoparticles and their integration in photonic circuitry will provide a step change in the already broad range of near-field directionality applications which are currently based on circular dipoles exclusively. We expect novel ideas to emerge in quantum optics, photonic nano-routing, photonic logical circuits, optical forces and torques of particles in near-field environments, inverse and reciprocal scenarios for polarization synthesis, integrated polarimeters, and other unforeseen devices throughout the whole electromagnetic spectrum.

\section{Acknowledgements}
This work was supported by European Research Council Starting Grant ERC-2016-STG-714151-PSINFONI and EPSRC (UK). A.Z. acknowledges support from the Royal Society and the Wolfson Foundation. All data supporting this research is provided in full in the main text.

\bibliographystyle{unsrt}
\bibliography{huygens}
\end{document}


\title{Janus and Huygens' dipolar sources for near-field directionality}
\author{Michela F. Picardi}
\email[Corresponding author: ]{michela.picardi@kcl.ac.uk}
\affiliation{Department of Physics, King’s College London, London WC2R 2LS, UK}
\author{Anatoly Zayats}
\affiliation{Department of Physics, King’s College London, London WC2R 2LS, UK}
\author{Francisco J. Rodr\'{i}guez-Fortu\~{n}o}
\affiliation{Department of Physics, King’s College London, London WC2R 2LS, UK}



\maketitle

\subsection{Deriving directionality from the angular spectrum of electric and magnetic dipole fields}

In the main text, we derived the condition for directionality using Fermi's golden rule. Here, we show that the same result can be obtained exploiting the angular spectrum representation of electric and magnetic dipoles sources.\\
We start from our results in Ref. \cite{picardi2017unidirectional} where we derived the angular spectra amplitudes of the electric field generated by electric and magnetic dipoles as:
\begin{eqnarray}
\mathbf{E}^{\mathrm{ED}}(k_x,k_y)|_{z=z_0}=\frac{i k^2}{8\pi^2\varepsilon}\frac{1}{k_z}\left[(\hat{\mathbf{e}}_s\cdot\mathbf{p})\hat{\mathbf{e}}_s+(\hat{\mathbf{e}}_p\cdot\mathbf{p})\hat{\mathbf{e}}_p\right],\nonumber\\
\mathbf{E}^{\mathrm{MD}}(k_x,k_y)|_{z=z_0}=-\frac{i k^2}{8\pi^2\varepsilon}\frac{1}{k_z}\frac{1}{c}\left[(\hat{\mathbf{e}}_p\cdot\mathbf{m})\hat{\mathbf{e}}_s-(\hat{\mathbf{e}}_s\cdot\mathbf{m})\hat{\mathbf{e}}_p\right]\nonumber,
\end{eqnarray}
these can be summed to give:
\begin{equation}
\mathbf{E}(k_x,k_y)|_{z=z_0}=\frac{i k^2}{8\pi^2\varepsilon}\frac{1}{k_z}\left[\left(\hat{\mathbf{e}}_s\cdot\mathbf{p}-\hat{\mathbf{e}}_p\cdot\frac{\mathbf{m}}{c}\right)\hat{\mathbf{e}}_s+\left(\hat{\mathbf{e}}_p\cdot\mathbf{p}+\hat{\mathbf{e}}_s\cdot\frac{\mathbf{m}}{c}\right)\hat{\mathbf{e}}_p\right].\label{eq:sumfield}
\end{equation}

The unit vectors $\hat{\mathbf{e}}_s$ and $\hat{\mathbf{e}}_p$ are the ones corresponding to $s$- and $p$-polarized fields, respectively, and are given by:
\begin{eqnarray}
\hat{\mathbf{e}}_s &=& \frac{\hat{\mathbf{z}}\times \hat{\mathbf{k}}}{\sqrt{(\hat{\mathbf{z}}\times \hat{\mathbf{k}})\cdot(\hat{\mathbf{z}}\times \hat{\mathbf{k}})}},\label{eq:es}\nonumber\\
\hat{\mathbf{e}}_p &=& \hat{\mathbf{e}}_s\times \hat{\mathbf{k}},\label{eq:ep}
\end{eqnarray}
\noindent where $\hat{\mathbf{k}} = \mathbf{k}/k$ is the normalized wave-vector, with the property that $\hat{\mathbf{k}}\cdot\hat{\mathbf{k}} = \hat{\mathbf{e}}_s\cdot\hat{\mathbf{e}}_s = \hat{\mathbf{e}}_p\cdot\hat{\mathbf{e}}_p = 1$ and $\hat{\mathbf{e}}_s\cdot\hat{\mathbf{e}}_p = 0$. Using Eq.~\ref{eq:ep}, we can then substitute $\hat{\mathbf{e}}_p$ into Eq.~\ref{eq:sumfield}, obtaining:
\begin{equation}
\mathbf{E}(k_x,k_y)|_{z=z_0}=\frac{i k^2}{8\pi^2\varepsilon}\frac{1}{k_z}\left\{ \left[\hat{\mathbf{e}}_s\cdot\mathbf{p}-(\hat{\mathbf{e}}_s\times \hat{\mathbf{k}})\cdot\frac{\mathbf{m}}{c}\right]\hat{\mathbf{e}}_s+\left[(\hat{\mathbf{e}}_s\times \hat{\mathbf{k}})\cdot\mathbf{p}+\hat{\mathbf{e}}_s\cdot\frac{\mathbf{m}}{c}\right]\hat{\mathbf{e}}_p\right\}.\label{eq:sumfield2}
\end{equation}

Applying the scalar triple product cyclic property, we can finally arrive at a compact exact mathematical expression that describes any arbitrary dipole source angular spectrum:
\begin{equation}\label{eq:efield}
\mathbf{E}(k_x,k_y)|_{z=z_0}=\frac{i k^2}{8\pi^2\varepsilon} \frac{1}{k_z}\left\{  \left[\hat{\mathbf{e}}_s\cdot\left(\mathbf{\hat{k}}\times \mathbf{p}+\frac{\mathbf{m}}{c}\right)\right]\hat{\mathbf{e}}_p+
\left[\hat{\mathbf{e}}_s\cdot\left(\mathbf{p}-\mathbf{\hat{k}}\times\frac{\mathbf{m}}{c}\right)\right]\hat{\mathbf{e}}_s\right\},
\end{equation}
from which the directional properties of any dipole source immediately follow, both in propagating and evanescent components. Eq.~\ref{eq:efield} is plotted in the main text for different sources, clearly showing the directionality properties.

Equation \ref{eq:efield} describes the fields of the isolated source in a homogeneous medium. We would like to consider the effects of placing this dipole in near proximity to a waveguide. Let's consider, for simplicity, a planar waveguide along the $xy$ plane, supporting a guided mode with propagation constant $k_m$. The fraction of power generated by the source that is coupled to this mode on each direction depends on the overlap between the source's angular spectra and the waveguide mode angular spectra \cite{teich1991fundamentals,picardi2017unidirectional}. The angular distribution of excitation of the mode in the waveguide will thus be proportional to the amplitude of the spectral component having the transverse wavevector $\mathbf{k}_t = (k_m \cos \phi$, $k_m \sin \phi)$ at each angle $\phi$. To determine the contrast ratio between light coupled in two opposite directions, we can reorient our axes such that $k_y=0$ without loss of generality, and compare the $\mathbf{k}_t=(-k_m,0)$ and $(+k_m,0)$ dipole components. In particular, we can make one of them identically zero. To achieve this, we simply have to equate the field angular spectrum (Eq.~\ref{eq:efield}) to zero at a certain $\hat{\mathbf{k}}$, i.e. forcing an evanescent component to be zero.  We only need to do this for the term in Eq.~\ref{eq:efield} corresponding to the same polarization as the mode we are interested in. For example, if the mode is $p$-polarized, we take the term polarized along $\hat{\mathbf{e}}_p$ and equate it to zero:
\begin{equation}
\hat{\mathbf{y}} \cdot \left(\mathbf{\hat{k}}\times \mathbf{p}+\frac{\mathbf{m}}{c}\right)=0,
\label{eq:ppolcondition}
\end{equation}
where we used the fact that $\hat{\mathbf{e}}_s = \hat{\mathbf{y}}$ when $k_y = 0$. Notice that by substituting $\hat{\mathbf{k}} = (1/k)(\pm k_m, 0, \pm i \alpha_m)$, Eq.~\ref{eq:ppolcondition} can be written as $p_x(\pm i \alpha_m/k) - p_z(\pm k_m/k) + m_y/c = 0$ which is exactly the same condition as in the main text when $k_m$ is real. In contrast to the equation derived from Fermi's golden rule, this time the positive (negative) sign in $\alpha_m$ corresponds to the fields above (below) the dipole, because $\hat{\mathbf{k}}$ refers to the wave-vector of the dipole fields, and not those of the waveguide mode.
\begin{figure}
\includegraphics[width=10cm]{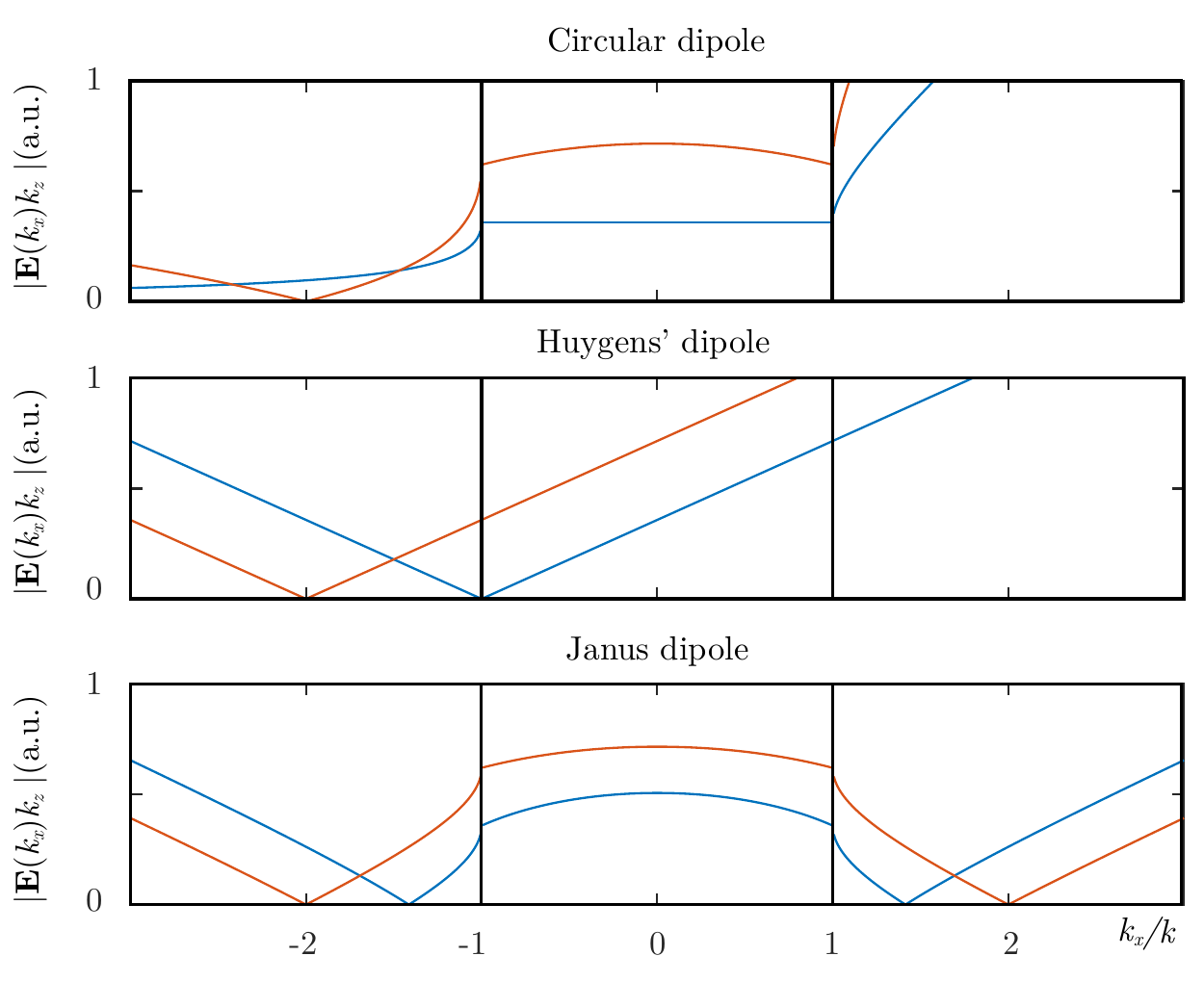}
\caption{Angular spectra of (blue) non optimized dipoles, compared with (red) optimized ones for $k_m=-2 k$. The non optimized dipoles are given by $(p_x,m_y/c,p_z) \propto (1,0,\pm i)$, $(0,\pm 1, 1)$ and $(1,\pm i,0)$ for circular, Huygens' and Janus respectively, while their optimized versions can be found in Table 1 in the main text.\label{fig:1}}
\end{figure}

\subsection{Alternative expression for the angular spectrum}

Curiously, substituting $\hat{\mathbf{e}}_p=\hat{\mathbf{e}}_s\times \hat{\mathbf{k}}$, in the $s$-polarized component of Eq.~\ref{eq:sumfield}, and $\hat{\mathbf{e}}_s=-\hat{\mathbf{e}}_p\times \hat{\mathbf{k}}$ in the $p$-polarized one, the two components can be written in a very convenient way:

\begin{eqnarray}
\mathbf{E}(k_x,k_y)|_{z=z_0}&=&\frac{i k^2}{8\pi^2\varepsilon}\frac{1}{k_z}\left\{ \left[\hat{\mathbf{e}}_s\cdot\mathbf{p}-(\hat{\mathbf{e}}_s\times \hat{\mathbf{k}})\cdot\frac{\mathbf{m}}{c}\right]\hat{\mathbf{e}}_s+\left[\hat{\mathbf{e}}_p\cdot\mathbf{p}-(\hat{\mathbf{e}}_p\times\hat{\mathbf{k}})\cdot\frac{\mathbf{m}}{c}\right]\hat{\mathbf{e}}_p\right\}\nonumber\\
&=&\frac{i k^2}{8\pi^2\varepsilon}\frac{1}{k_z}\left\{\left[ \hat{\mathbf{e}}_s\cdot\left(\mathbf{p}-\hat{\mathbf{k}}\times\frac{\mathbf{m}}{c}\right)\right]\hat{\mathbf{e}}_s+\left[\hat{\mathbf{e}}_p\cdot\left(\mathbf{p}-\hat{\mathbf{k}}\times\frac{\mathbf{m}}{c}\right)\right]\hat{\mathbf{e}}_p\right\}\nonumber\\
&=&\frac{i k^2}{8\pi^2\varepsilon}\frac{1}{k_z}\left(\mathbf{p}-\hat{\mathbf{k}}\times\frac{\mathbf{m}}{c}\right)\left[\left(\cdot\ \hat{\mathbf{e}}_s \right)\hat{\mathbf{e}}_s + \left(\cdot\ \hat{\mathbf{e}}_p\right)\hat{\mathbf{e}}_p\right].\nonumber
\end{eqnarray}
In this way, the amplitude of the $s$-polarized component is given by the dot product of $\hat{\mathbf{e}}_s$ with the vector $\mathbf{v}=\mathbf{p}-\left(\hat{\mathbf{k}}\times\frac{\mathbf{m}}{c}\right)$, and the amplitude of the $p$-polarized component is given by the dot product of $\hat{\mathbf{e}}_p$ with the same vector $\mathbf{v}$. This is even more compact than Eq.~\ref{eq:efield} but is slightly less convenient for the derivation of optimized dipoles.

\bibliographystyle{apsrev}
\bibliography{huygens.bib}